\documentclass[a4paper,12pt]{article}  
\usepackage[dvips]{graphicx}
\usepackage{amssymb}

\newcommand{\half}{{\textstyle{1\over 2}}} 
 
\newcommand{\goes}{\rightarrow} 
 
\newcommand{\GeV}{\; \mathrm{GeV}} 
\newcommand{\TeV}{\; \mathrm{TeV}} 
\newcommand{\beq}{\begin{equation}} 
\newcommand{\eeq}{\end{equation}} 
\newcommand{\bea}{\begin{eqnarray}} 
\newcommand{\eea}{\end{eqnarray}}
\newcommand{\etal}{\textit{et. al.}}

\newcommand{\strahlo}{e^+e^- \goes Z h^0 \goes \bar{\nu} \nu \, h^0}
\newcommand{\wwfuso}{e^+e^- \goes \bar{\nu}_e \nu_e \,  W W 
                                  \goes \bar{\nu}_e \nu_e \,  h^0}

\newcommand{\higproo}{e^+e^- \goes \bar{\nu} \nu h^0/H^0}

\newcommand{\strahl}{e^+e^- \goes Z h^0 \goes \bar{\nu} \nu \, H^0_k}

\newcommand{\higpro}{e^+e^- \goes \bar{\nu} \nu H^0_k}

\def\polm{{\cal P}_{-}}
\def\polp{{\cal P}_{+}}
\def\polLR{{\cal P}_{LR}}
\def\polRL{{\cal P}_{RL}}

\def\e{\epsilon}

\def\nn{\nonumber}

\def\sq{{\tilde q}}
\def\st{{\tilde t}}
\def\sb{{\tilde b}}

\begin{document} 
\begin{titlepage} 
 
\begin{flushright}
HEPHY-PUB 763/02 \\  
hep-ph/0210038  
\end{flushright} 
\begin{center} 
\vspace*{1.5cm} 
 
{\Large{\textbf {Single Higgs boson production at future linear colliders
including radiative corrections}}}\\ 
\vspace*{10mm} 
 
{\large H.~Eberl, W.~Majerotto, and V.~C.~Spanos} \\ 

\vspace{.7cm} 

{\it Institut f\"ur Hochenergiephysik der \"Osterreichischen Akademie
der Wissenschaften,}\\
{\it A--1050 Vienna, Austria}

\end{center} 
\vspace{3.cm} 
\begin{abstract} 
The next generation of high energy $e^+ e^-$ linear colliders 
is expected to operate at $\sqrt{s} \gtrsim 500 \GeV$.
In this energy range the $WW$ fusion
channel dominates the Higgs boson production 
cross section $e^+ e^- \goes \bar{\nu} \nu h^0/H^0$. 
We calculate the one-loop corrections to this
process due to fermion and sfermion loops within the MSSM. 
We perform a detailed numerical analysis of the 
total cross section and the distributions of the rapidity,
the transverse momentum and the production angle of the Higgs boson.
The  fermion-sfermion correction is substantial  being
of the order of $-10\%$ and  is dominated by the fermion loops. 
In addition, we explore the possibility of polarized $e^+ / e^-$ beams.
In the so-called ``intense coupling'' scenario the
production of the heavy Higgs boson $H^0$ is also discussed.

\end{abstract} 
\end{titlepage} 

\baselineskip=18pt
\section{Introduction}
The Standard Model (SM) of fundamental particles has been tested
with an impressive precision by a large number of experiments.
The resulting body of data is consistent with the matter content
and gauge interactions of the SM and a Higgs boson $h^0$ of mass
$m_{h^0} \leq 204 \GeV$ \cite{all}. The four  experiments
at LEP delivered a lower bound for the SM Higgs boson mass,
$m_{h^0} \gtrsim 114 \GeV$ \cite{higgs}.
If a fundamental Higgs boson exists, it would fit very naturally
into supersymmetric (SUSY) extensions of the SM, in particular into the
Minimal Superymmetric Standard Model (MSSM).
The latter requires the existence of two isodoublets of scalar Higgs
fields, implying three neutral Higgs bosons, two $CP$-even bosons
$h^0$, $H^0$, one $CP$-odd $A^0$, and two charged Higgs bosons $H^\pm$.
The lightest Higgs particle $h^0$ could exhibit properties
similar to those of the SM Higgs boson. Its mass is predicted
to be less than $135 \GeV$ \cite{higgslim}, taking into
account radiative corrections.
The present experimental bound from LEP are $m_{h^0} > 88.3 \GeV$
and $m_{A^0} > 88.4 \GeV$ at $95 \%$ CL \cite{higgs}.

The next step in the search for the Higgs boson will take place at the 
Tevatron \cite{tevatron,carhaber} in $p\, \bar{p}$ collisions at $2 \TeV$. 
The gluon-gluon fusion process is the dominant neutral Higgs production
mechanism, but suffers from the overwhelming QCD background of $b\, \bar{b}$
production. The most promising Higgs discovery mechanism for
$m_{h^0} < 130 \GeV$ is most likely the Higgsstrahlung
$q \, \bar{q} \goes W \goes W h^0$.
The $WW$ fusion process $WW \goes h^0$, i.e. $p\, \bar{p} \goes
q \, WW \, \bar{q} \goes q \, h^0 \, \bar{q}$, 
plays a less important r\^ole.

At LHC, in $p\, p$ collisions at $14 \TeV$, the gluon-gluon fusion
mechanism provides the dominant contribution to Higgs boson
production \cite{aachen}. 
The next important Higgs production channel is the vector 
boson fusion $VV \goes h^0 / H^0$. In particular, it provides an
additional event signature due to the two energetic forward
jets. It has been argued that the channels $WW \goes h^0/H^0 \goes
\tau \, \bar{\tau}$ and $WW$ can serve as suitable search channels
at LHC, even for a Higgs boson mass of $m_h \sim 120 \GeV$ \cite{zepe}.
Very recently, it has been shown that the $WW$ fusion process
$q\, q \goes q\, h^0 \, q $ with $h^0 \goes b \, \bar{b}$ may
be used to identify and study a light Higgs boson at the LHC due the two
rapidity gaps in the final state \cite{roeck}

The next generation of high energy $e^+ e^-$ linear colliders 
is expected to operate in the energy range of
$\sqrt{s}=300-1000 \GeV$ (JLC, NLC, TESLA) \cite{jlc,nlc,tesla}.
The possibility of a multi-TeV linear collider with
$\sqrt{s} \sim 3\TeV$ (CLIC) is also under study \cite{clic}.  
At these colliders high-precision analyses of the Higgs boson
will be possible. In $e^+ e^-$ collision, for energies $\gtrsim 200\GeV$,
the production of a single Higgs boson plus missing energy 
starts to be dominated by $WW$ fusion \cite{fusion,alta,kilian}, that
is $\wwfuso /H^0$, whereas the Higgsstrahlung process \cite{ellis}
$\strahlo$ becomes less important. The rates for the $ZZ$ fusion are
generally one order of magnitude smaller than those of the $WW$ channel.

The process $\wwfuso /H^0$ was calculated at 
tree-level in Refs.~\cite{fusion,alta,kilian}. The leading one-loop 
corrections to the $WWh^0$ vertex in the SM were also calculated 
(see the review article \cite{kniehl-review} and the references therein).
For this coupling also QCD corrections were included, the 
$\mathcal{O}(a_s\,G_F\,m_t^2)$ corrections in Ref.~\cite{kniehl1} and
the  $\mathcal{O}(a_s^2\,G_F\,m_t^2)$ ones in Ref.~\cite{kniehl2}.

In this paper, we have calculated the  one-loop corrections to
the $WWh^0/H^0$ vertex in the MSSM due to  
fermion/sfermion loops. We have also included the corresponding wave-function
corrections to the $W$ and Higgs bosons.
They are supposed to be the dominant corrections due 
to the Yukawa couplings involved.
We have applied the corrections to the single Higgs boson production in
$e^+e^-$ annihilation in  the energy range $\sqrt{s}=0.5-3\TeV$, i.e. to
$\wwfuso /H^0$. 
We have also included the Higgsstrahlung 
process $\strahlo/H^0$ and the interference
between those two mechanisms. Because the Higgsstrahlung process is much
smaller in this range, we have neglected 
its radiative corrections~\cite{hollik}.

In a previous paper~\cite{EMS} we have already given some results
for the light Higgs boson production. This work represents
a much more detailed study of single Higgs boson ($h^0$ and $H^0$)
production in $e^+ e^-$ collisions, including radiative corrections.
In particular, important kinematical distributions of the
rapidity, transverse momentum and the production angle of the
Higgs boson are given. Polarization of the incoming  $e^+/e^-$
beams is also included. 
In addition, we consider the ``intense coupling regime''\cite{intense}, where
all Higgs bosons of the MSSM are rather light, and for large
$\tan\beta$ couple maximally to electroweak gauge bosons and strongly
to the third generation fermions. The paper  contains
a brief discussion on the background, although it has not been 
our intention to perform a Monte Carlo study.

The paper is organized as follows: In section~\ref{sec:loop} 
we give the formulae 
for the tree-level amplitude, then calculate the one-loop
corrections, taking into account fermion/sfermion loops.
We express the corrections to the vertex in terms of form factors.
In section~\ref{sec:xsec}  we are discussing the calculation of the 
cross section for the Higgs boson production, especially including
the one-loop correction.
 In section~\ref{sec:results}, we perform a detailed numerical analysis
and discuss our results. 
Finally, in section~\ref{sec:concl}  we present our conclusions.
Appendix~\ref{sec:app_ff}  exhibits explicitly 
the expressions for the form factors.
Appendix~\ref{sec:app_xsec}  gives details of 
the calculation of the cross section and
the various distributions.

\section{Matrix elements and  one-loop corrections}
\label{sec:loop}
We study the process
\begin{equation}
e^+(p_2) + e^-(p_1) \to H^0_k(p) + \bar\nu(p_4) + \nu(p_3)\,.
\label{eq:LCsingleHiggs}
\end{equation}
$p$, $p_1$, $p_2$, $p_3$, $p_4$ are the corresponding
four momenta and
 $H^0_k = \{h^0,\, H^0\}$.

\begin{figure}[t]  
\centering\includegraphics[scale=1.]{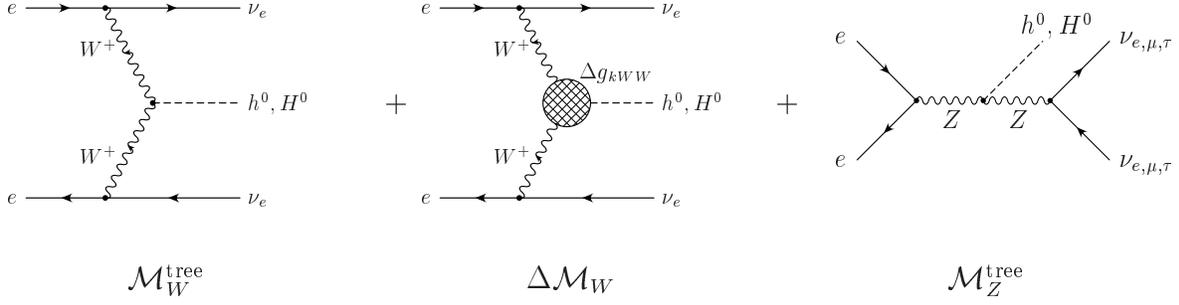} 
\caption[]{ The Feynman graphs for the process $\higproo$.
For the Higgsstrahlung contribution 
 $|{\cal M}_Z^{\rm tree}|^2$ one has to sum over all
three neutrino types.} 
\label{fig:corr}
\end{figure}

The contributing Feynman graphs are shown in Fig.~\ref{fig:corr}.
The amplitude of Eq.~(\ref{eq:LCsingleHiggs}) consists of three parts:
$WW$ fusion at tree level ${\cal M}_W^{\rm tree}$,
its one-loop correction  $\Delta {\cal M}_W$ and the Higgsstrahlung
process ${\cal M}_Z^{\rm tree}$, $\strahl$, i.e. 
${\cal M} = {\cal M}_W^{\rm tree} 
 + {\cal M}_Z^{\rm tree} + \Delta {\cal M}_W$.
As already mentioned in the introduction, we have neglected the
radiative corrections to ${\cal M}_Z^{\rm tree}$,
as this amplitude is much smaller than ${\cal M}_W^{\rm tree}$
in the energy region considered.
We will include polarization of the incoming electron and
positron beams.
$\polm$ and $\polp$ denote 
the polarization of the $e^-$ and $e^+$ beams, with
the convention ${\cal P}_{\pm} = \{-1, 0, +1\}$ for $\{$left-polarized,
unpolarized, right-polarized$\}$ $e^\pm$ beams, respectively. (e.~g.,
$\polm = - 0.8$ means that 80\% of the electrons are left-polarized and 
20\%  unpolarized.) The mass of the electron is negligible. 
Therefore, all vector particles propagate only transversally. 
We introduce the polarisation factors
\begin{equation}
  \polLR = (1 - \polm)\, (1 + \polp)\, , \qquad
  \polRL = (1 + \polm)\, (1 - \polp)\, .
\label{eq:polLR}
\end{equation}
$\polLR \; (\polRL)$ gets maximal 
in $e^-_L\,e^+_R \; (e^-_R\,e^+_L)$ collisions.

The squared matrix element in the 
one-loop approximation is  given by 
\begin{equation}
 |{\cal M}|^2 =|{\cal M}_W^{\rm tree}|^2 + |{\cal M}_Z^{\rm tree}|^2 +
  2\, \Re\left[{\cal M}_Z^{\rm tree}({\cal M}_W^{\rm tree})^\dagger  +
  \Delta {\cal M}_W \left(({\cal M}_W^{\rm tree})^\dagger +
  ({\cal M}_Z^{\rm tree})^\dagger\right)\right] \,.
\label{eq:matsquared}
\end{equation}
The first three terms correspond to the tree level part \cite{kilian}:
\begin{eqnarray}
|{\cal M}_W^{\rm tree}|^2 & = &  \polLR\, g^2_{k W W}\, g^4 \,
   \frac{1}{(k_1^2 - m_W^2)^2\, (k_2^2 - m_W^2)^2}\,
    p_1.p_4\, p_2.p_3\,,
\label{eq:matWtreesquared}\\
|{\cal M}_Z^{\rm tree}|^2 & = & g^2_{k Z Z}\,  \frac{g^4}{c_W^4} \,
\frac{1}{(q_1^2 - m_Z^2)^2\,
\left((q_2^2 - m_Z^2)^2  + m_Z^2\,\Gamma_Z^2\right)}\, \times \nonumber\\[2mm]
&& \hspace{1cm}
    \left(\polLR\, (C_L^e)^2\, p_1.p_4\, p_2.p_3 +
\polRL\, (C_R^e)^2\, p_1.p_3\, p_2.p_4 
\right) \,,
\label{eq:matZtreesquared}\\
  2\, \Re\left[{\cal M}_Z^{\rm tree}({\cal M}_W^{\rm tree})^\dagger\right] 
 & = &
  2\, \polLR\, \, C_L^e\, g_{k Z Z}\, g_{k W W}\, \frac{g^4}{c_W^2}
\,\, p_1.p_4\, p_2.p_3\,\times\nonumber\\
&& \hspace{-0.5cm}
   \frac{q_2^2 - m_Z^2}{(q_1^2 - m_Z^2)\,
\left((q_2^2 - m_Z^2)^2  + m_Z^2\,\Gamma_Z^2\right)(k_1^2 - m_W^2)
(k_2^2 - m_W^2)}\, ,
 \label{eq:matZWtree}
\end{eqnarray}
$k = 1,2$, where 1 (2) stands for $h^0$ ($H^0$).

The $W W H^0_k$ couplings are
$g_{1 W W} = g\, m_W \sin(\beta - \alpha)$, 
$g_{2 W W} = g\, m_W \cos(\beta - \alpha)$, 
$g_{1 Z Z} = \frac{g}{c_W}\, m_Z \sin(\beta - \alpha)$, 
$g_{2 Z Z} = \frac{g}{c_W}\, m_Z \cos(\beta - \alpha)$, 
$\beta = \arctan(v_2/v_1)$, $\alpha$ is the $h^0$--$H^0$ mixing angle,
$k_1 = p_3 - p_1$, $k_2 = p_2 - p_4$, $q_1 = p_1 + p_2$, $q_2 = p_3 + p_4$,
$\Gamma_Z$ is the total $Z$-boson width, $C_L^e = \sin^2\theta_W - 1/2$, 
$C_R^e = \sin^2\theta_W$, with $\theta_W$ being the Weinberg angle,
and $c_W \equiv \cos\theta_W$.

Notice that the $WW$ fusion is enhanced if the electron has
left and positron right polarization. For instance,
with $\polm=-0.85$, $\polp=0.6$, one has $\polLR=2.96$ and
$\polRL=0.06$. We are of course interested in the case where the fusion
process dominates over the Higgsstrahlung process
to get a large Higgs production rate.
In this case the polarized cross section is just given by the
unpolarized one times $\polLR$.

Now we turn on discussing the calculation of the one-loop correction due  
to  the fermion and sfermion  loops.
One expects them to be the most important corrections due
to the Yukawa couplings involved.
The renormalization of the five-point function
simplifies to the renormalization of the $W W H^0_k $  vertex with 
off-shell vector bosons, 
where the renormalization of the other two vertices in 
the process (e.~g.  the $e^- \nu_e W^+$ coupling) is absorbed. 
The contributions of the first and second families of (s)fermions
are numerically negligible
due to the smallness of their Yukawa couplings.
Therefore, we will consider the contribution arising from
the third family of (s)fermions.

\begin{figure}[th]  
\centering\includegraphics[scale=.85]{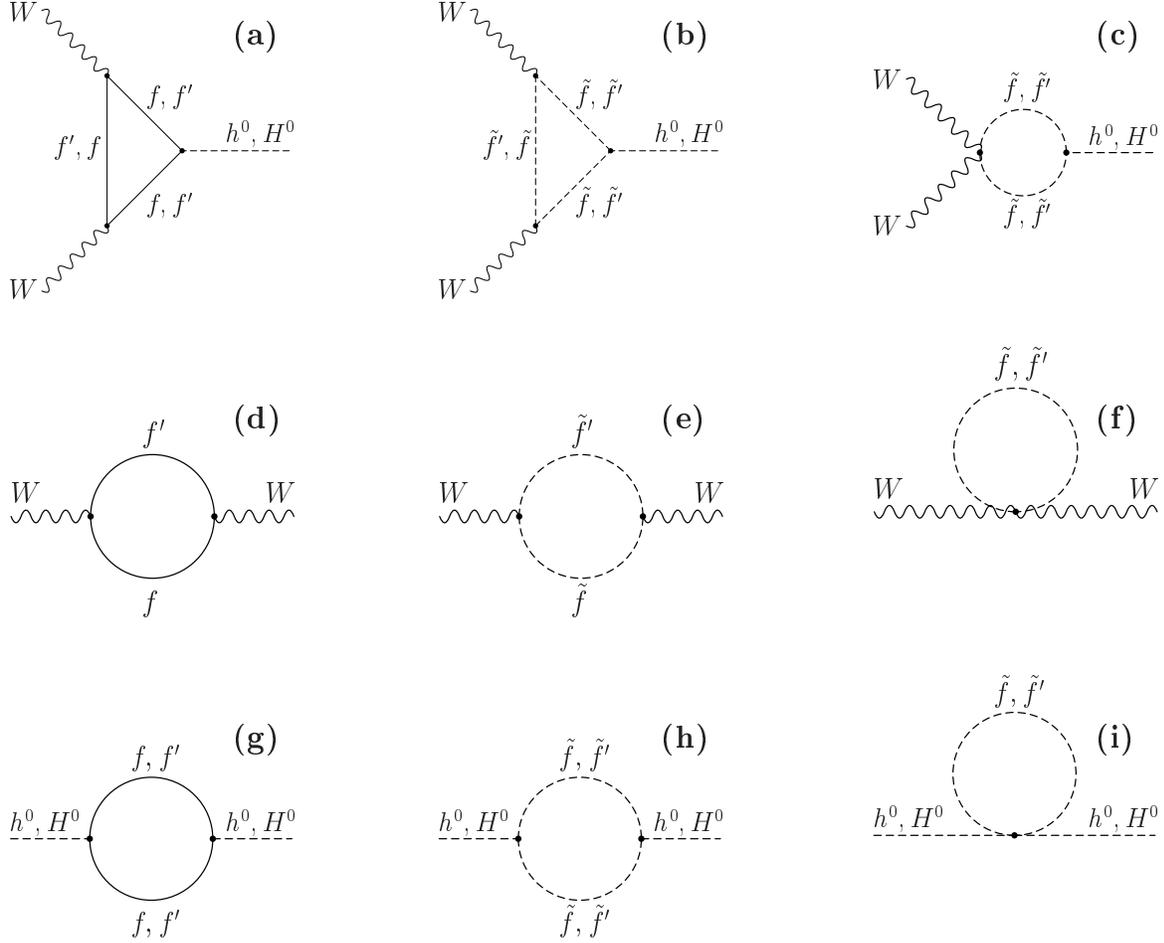}  
\caption[]{The Feynman graphs that contribute to the vertex 
(a)--(c)  and the wave-function corrections (d)--(i). 
 $f$ ($f'$) denotes the up (down) type fermion.} 
\label{fig:feyn}
\end{figure}

At the one-loop level the Lagrangian for the  
$W W H^0_k$ coupling can be written as
\beq
{\mathcal L}=\left( g_{k W W} \,g^{\mu \nu}
+N_c\,\left(\Delta g_{k W W}\right)^{\raisebox{-1.3mm}{$\scriptstyle \mu\nu$}}
 \right) \, 
       H^0_k \, W^+_\mu \, W^-_{\nu}\, .
\label{eq:lagr}
\eeq
The colour factor $N_c$ is 3 for (s)quarks and 1 for (s)leptons.
Actually, for the calculation of the one-loop corrected 
$W W H^0_k$ vertex  one has to compute
the vertex and the wave-function
corrections due to the graphs of Fig.~\ref{fig:feyn}, as
well as the coupling correction  $\delta g_{k WW}^{(c)}$, 
\beq
 \left(\Delta g_{k WW}\right)^{\raisebox{-1.3mm}{$\scriptstyle \mu\nu$}} =
 \left(\delta g_{k WW}^{(v)}\right)^{\raisebox{-1.3mm}{$\scriptstyle \mu\nu$}}
 +
\left(\delta g_{k WW}^{(w)} 
   +  \delta g_{k WW}^{(c)}\right) g^{\mu\nu} \, .
\label{eq:ct}
\eeq
The vertex correction can be
expressed in terms of  all possible form factors,
\beq
  \left(\delta g_{k WW}^{(v)}\right)^{\raisebox{-1.3mm}{$\scriptstyle \mu\nu$}}
 =
   F^{00} g^{\mu\nu} + F^{11} k_1^\mu k_1^\nu 
 + F^{22} k_2^\mu k_2^\nu 
 + F^{12} k_1^\mu k_2^\nu 
 + F^{21} k_2^\mu k_1^\nu  
 + i\,F^\epsilon \epsilon^{\mu\nu\rho\delta} 
                         k_{1\rho} k_{2\delta} \, ,
\label{eq:vertex}
\eeq
$k_{1,2}$ denote the four-momenta 
of the off-shell $W$-bosons. 
At tree-level only the 
structure with $g_{\mu \nu}$ is present, and therefore
all form factors but $F^{00}$ have to be 
ultra violet (UV) finite without being renormalized.

The  wave-function correction is 
\begin{equation}
\delta g_{k WW}^{(w)} = g_{kWW}\, \left( \half \, (\delta Z_H)_{kk} 
 + \delta  Z_W\right) + \half \, g_{lWW}\, (\delta Z_H)_{l k}  \, ,
\label{eq:wave}
\end{equation}
$l \ne k$, and $\delta Z_H$ and $\delta  Z_W$ 
are the symmetrized Higgs boson and the \mbox{$W$-boson} wave-function
corrections calculated from the graphs (g)--(i) and (d)--(f)
of the Fig.~\ref{fig:feyn},  respectively. 

In the case of the off-shell $W$-bosons coupling to $e\,\nu_e$, $\delta  Z_W$ has the form
\beq
\delta Z_W
   = \frac{\delta m_W^2 - \Re \Pi_{WW}^T (k_1^2)}{k_1^2 - m_W^2} 
+ \frac{\delta m_W^2 - \Re \Pi_{WW}^T (k_2^2)}{k_2^2 - m_W^2} 
+ 2\, \frac{\delta g}{g}\, .
\label{eq:dzw}
\eeq
The  coupling correction is 
\beq
\delta g_{k WW}^{(c)}  =  
\left( \frac{\delta g}{g} + \frac{\delta m_W}{m_W}\right) g_{k WW} 
           - (-1)^k\, \frac{\sin2\beta}{2}\,
\frac{\delta \tan\beta}{\tan\beta} \, g_{l WW}\, , 
\label{eq:dgc}
\eeq
$l \ne k$.
The expressions on the right-hand 
sides of the Eqs.~(\ref{eq:wave})--(\ref{eq:dgc}) 
can be found in Ref.~\cite{EMKY}.
Especially, we fixed the counter term $\delta\tan\beta$ by the
on-shell condition $\Im \, \hat{\Pi}_{A Z}(m^2_A)=0$,
where  $\hat{\Pi}_{A Z}(m^2_A)$ is the renormalized self-energy
for the mixing of the pseudo-scalar Higgs boson $A^0$ and $Z$-boson.
By adding  the vertex correction, Eq.~(\ref{eq:vertex}), the wave-function,
Eq.~(\ref{eq:wave}), and coupling correction, Eq.~(\ref{eq:dgc}), 
we get the renormalized and therefore UV finite one-loop correction
\beq
\left(\Delta g_{k W W}\right)^{\raisebox{-1.3mm}{$\scriptstyle \mu\nu$}}  =
  \hat  F^{00} g^{\mu\nu} + F^{11} k_1^\mu k_1^\nu 
 + F^{22} k_2^\mu k_2^\nu 
 + F^{12} k_1^\mu k_2^\nu 
 + F^{21} k_2^\mu k_1^\nu  
 + i\,F^\epsilon \epsilon^{\mu\nu\rho\delta} 
                         k_{1\rho} k_{2\delta} \, ,
\label{eq:giWWoneloop}
\eeq
which has exactly the same form as Eq.~(\ref{eq:vertex}) but the form factor $F^{00}$ is 
substituted by the renormalized and hence UV finite one,
\beq
\hat F^{00} = F^{00} + \delta g_{k WW}^{(w)} + \delta g_{k WW}^{(c)}\, .
\eeq

Having calculated the form factors of  Eq.~(\ref{eq:giWWoneloop}),
one can proceed to the calculation of the one-loop corrected
cross section. 
The remaining parts of  Eq.~(\ref{eq:matsquared})
due to the one-loop corrections are
\bea
  2 \Re\left[\Delta {\cal M}_W 
     \left({\cal M}_W^{\rm tree}\right)^\dagger\right] &\! =\! &  
  \polLR\; g_{k W W} \; g^4\; 
 \left(2\,\hat F^{00}\, p_1 \cdot p_4 \, p_2\cdot p_3 
 + F^{21}\, S\right) \nonumber\\
&& \times  \prod_{i=1,2} \frac{1}{(k_i^2 - m_W^2)^2}\, , 
\label{eq:wwcor}\\ 
2 \Re\left[\Delta {\cal M}_W
 \left({\cal M}_Z^{\rm tree}\right)^\dagger\right] 
&\!=\!& \polLR\; C_L^e\, g_{k Z Z}\;  \frac{g^4}{c^2_W}
\left(2\,\hat F^{00}\, p_1\cdot p_4\, p_2\cdot p_3 + F^{21}\, S\right)\, 
    \nonumber\\
&& \times \frac{q_2^2 - m_Z^2}{(q_1^2 - m_Z^2)\,
\left((q_2^2 - m_Z^2)^2  + m_Z^2\,\Gamma_Z^2\right)}
\prod_{i=1,2} \frac{1}{k_i^2 - m_W^2}
\, ,
\label{eq:zzcor} 
\eea
where
\bea
S &=& (p_1\cdot p_4 + p_2\cdot p_3)
  \left( p_1\cdot p_2\, p_3\cdot p_4 +
      p_1\cdot p_4\, p_2\cdot p_3  - p_1\cdot p_3\, p_2\cdot p_4\right)
 \nonumber\\
  &-& 2\,(p_1\cdot p_2 + p_3\cdot p_4)\,p_1\cdot p_4\, p_2\cdot p_3 \, .
\label{eq:sfact}
\eea
As $k_{1,2}$ are spacelike,  $F^{00}$ and $F^{21}$ have no 
absorptive parts and are therefore real.
The term with  $F^\e$ does not contribute to the cross section. 
The explicit forms of $F^{00}$ 
and $F^{21}$ are given in  Appendix~\ref{sec:app_ff}.

\section{Calculation of the cross section}
\label{sec:xsec}
In order to calculate the cross section, including the radiative
corrections from fermion and sfermion loops, one has to choose an
appropriate reference frame. A detailed discussion on
this is given in Appendix~\ref{sec:app_xsec}.
The momenta of the particles participating in the process
are defined in Eq.~(\ref{eq:LCsingleHiggs}).

\begin{figure}[t] 
\centering\includegraphics[scale=.55]{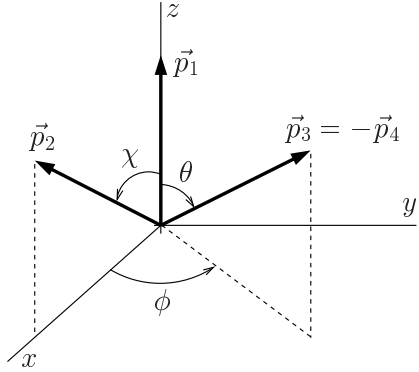}
\caption[]{ The integration variables in the rest  
frame of the two final fermions.}
\label{frame34} 
\end{figure}  

The calculation of the total cross section for the Higgs boson
production $\higpro$ is performed in two steps. 
First, we calculate the differential cross section
\beq
E_p \,\frac{d^3 \sigma}{d^3 p}=
           \int \frac{|{\mathcal{M}}|^2}{16 \, s \, (2\pi)^5}
  \;  \delta^4(p_1 + p_2 - p_3 - p_4 -p) \; 
\frac{d^3p_3}{E_3} \; \frac{d^3p_4}{E_4} \, ,
\label{eq:gxsecm}
\eeq
where $E_p$ is the energy of the produced Higgs boson. 
For this calculation it is convenient
to work in the rest frame of the two final fermions, where
one has $\vec{p}_3+\vec{p}_4=0$, see Fig.~\ref{frame34}. 
In this frame the differential
cross section can be evaluated using 
\beq
E_p \, \frac{d^3 \sigma}{d^3 p}
 = \int_{-1}^{1}\, d \cos\theta \int_{0}^{2\pi} \, d \phi \;
 \frac{|{\mathcal{M}}|^2}{ s \, (4\pi)^5} \, ,
\label{eq:xsec1m}
\eeq
by integrating the
total amplitude $|\mathcal{M}|^2$
over the angles $\theta$, $\phi$, as they are defined in Fig.~\ref{frame34}. 
For the tree-level case, it has been shown \cite{alta,kilian} that
these integrations can be performed analytically.
For example, the results for the fusion process, the Higgsstrahlung, and
their interference term from 
Eqs.~(\ref{eq:matWtreesquared})--(\ref{eq:matZWtree})
can be found in Eqs.~(5)--(8) of Ref.~\cite{kilian}.
One major complication of the inclusion of the one-loop corrections
of Eqs.~(\ref{eq:wwcor}) and (\ref{eq:zzcor}) is 
that it is not possible anymore to
calculate these integrals analytically. This is due to the
fact that the form factors $\hat{F}^{00}$ and  $F^{21}$ are functions
of the momentum transfer $k_{1,2}^2$.     
Therefore, for the one-loop corrected cross section we are bound 
to use numerical methods for this task.

\begin{figure}[t] 
\centering\includegraphics[scale=.57]{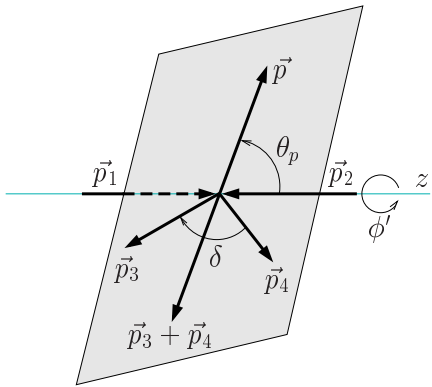}
\caption[]{The momenta in the the rest  
 frame of the two initial fermions.}
\label{frame12} 
\end{figure}

The second step consists of the integration of the differential cross section
in order to get the total  for the Higgs boson production.
To do this, we are working in the rest frame of the initial fermions,
where $\vec{p}_1+\vec{p}_2=0$, see Fig.~\ref{frame12}. 
In this reference frame we obtain the total cross section
\beq
\sigma =  2 \pi \, \int_{-1}^1 d \cos\theta_p \,\,
\int_{m_H}^{E_p^{\mathrm{max}}} dE_p \,\,
  \sqrt{E_p^2 - m^2_{H_k^0}}\,  \left(  E_p \,\frac{d^3 \sigma}{d^3 p} \right) \,,
\label{eq:xsec2m}
\eeq
where $\theta_p$ denotes the  angle of the produced Higgs boson with
respect to the beam direction.
Alternatively, one can use the rapidity $y$ and the transverse 
momentum $p_T$ of the Higgs boson and   calculate
the cross section as
\beq
\sigma = \int_{y_{-}}^{y_{+}} dy 
         \int_0^{(p_T^2)^{\mathrm{max}}} d p_T^2 
     \left( \frac{d^2 \sigma} {dy \, dp_T^2} \right) \,,
\label{eq:xsec3m}
\eeq  
where the integrand and the integration
limits are given in Eqs.~(\ref{eq:relation}) and (\ref{eq:limits}),
respectively.
These integrations, even in the tree-level case, 
are carried out  numerically.
The advantage of using the $y$, $p_T$ variables is the faster 
numerical convergence of the integration routines, due to the
strong forward-backward peaking of the differential cross section
for large $\sqrt{s}$.
For the calculation of the one-loop corrected cross section, 
one has to perform four numerical integrations successively.
For this purpose, we use appropriate  
numerical integration routines found in the {\small NAG} library.
In addition, we have checked that for the tree-level case our
completely numerical calculation agrees with high
accuracy with the semi-analytical results of Ref.~\cite{kilian}.  
 
\section{Discussion and results}
\label{sec:results}
In our numerical analysis, we have taken into account the 
contribution arising from the third family of fermion/sfermion loops.
This contribution turns out to be the dominant one, in
comparison with the first two families corrections, due
to the large values of the Yukawa couplings $h_t$ and $h_b$.
The impact of the running of the electroweak couplings
$g$ and $g'$ is not negligible, especially for
large $\sqrt{s}$ we are discussing here. 
Therefore it  has been taken into account.
For the calculation of the SUSY  Higgs boson spectrum
and the Higgs mixing angle $\alpha$, a computer  
program based on  Ref.~\cite{carena} has been used. 
We note that for values of $\tan\beta > 5$ and $m_A$ large the
SUSY $W W h^0$ coupling mimics the SM one, while the
$W W H^0$ is very small. This is due to fact that
for these values of $\tan\beta$ we have $\sin(\beta-\alpha) \sim 1$
and $\cos(\beta-\alpha) \sim 0$.

In the so-called ``intense coupling regime'' \cite{intense},
where the neutral Higgs bosons are almost degenerate and light,
 $m_H \sim m_h \sim m_A \sim 100 \GeV$, there is the possibility
that the $WW h^0$ coupling is suppressed, while  the $W W H^0$ one is
{\em not} suppressed. For this case it is worth to
explore the possibility of the heavy Higgs production.

For simplicity, for all plots we have used
$A_t=A_b=A_\tau=A$, 
$\{m_{\tilde{U}},m_{\tilde{D}},m_{\tilde{L}},m_{\tilde{E}}\}=$
$\{ \frac{9}{10},\frac{11}{10},1,1 \} \, m_{\tilde{Q}} $
and  $M_1=\frac{5}{3}\, \tan^2 \theta_W\, M_2 \sim 0.5\,M_2$.
The choice of a common trilinear coupling and
the correlation between  the soft sfermion and gauginos masses 
are inspired by unification.

Concerning the polarization, 
from the Eqs.~(\ref{eq:matWtreesquared})--(\ref{eq:matZWtree}),(\ref{eq:wwcor})
and (\ref{eq:zzcor}), we can see that basically the polarized
cross section for the process $\higproo$ is the unpolarized  multiplied
with the factor $\polLR$. Although the
second term of the Higgsstrahlung contribution
 in Eq.~(\ref{eq:matZtreesquared})  is multiplied
with  $\polRL$, considering that for $\sqrt{s}>500 \GeV$ 
the total tree-level cross section is dominated by the $WW$ fusion
channel, this term is not important 
for the Higgs production at  future linear colliders. 
Therefore, if one wants to enhance the  cross section
the appropriate mode would be $e_L^- \, e_R^+$, where
we have  $\sigma_{\rm pol} \simeq \polLR \, \sigma_{\rm unpol}$,
while $\polLR$ being as high as 3 to 4.

\begin{figure}[t]  
\begin{center}
\includegraphics[scale=.75]{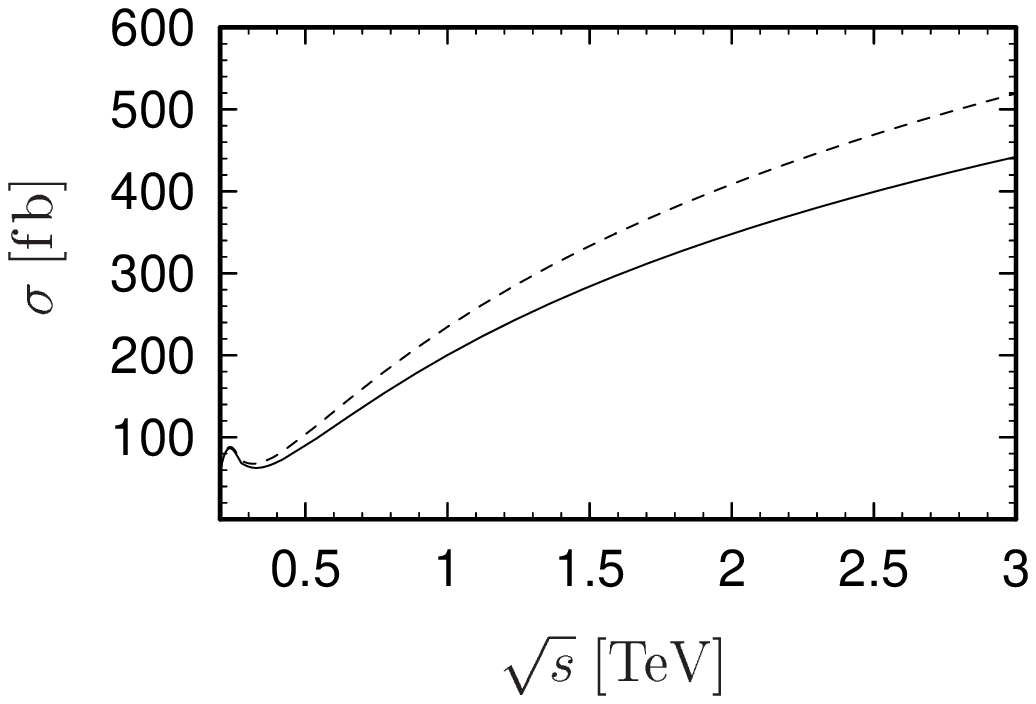} 
\includegraphics[scale=.75]{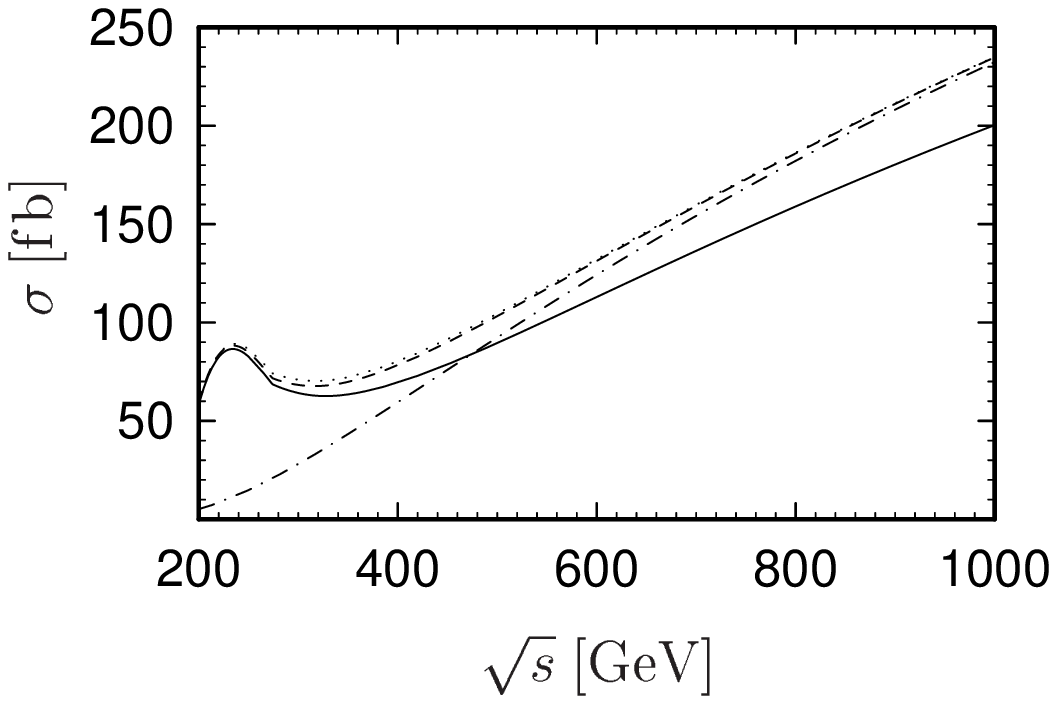}
\end{center}

\caption[]{The tree-level cross section $\sigma_0$ (dashed line) and
the one-loop corrected one $\sigma$ (solid line) for $\sqrt{s}$ up
to $3\TeV$ (left). The SUSY parameters are chosen as
$\tan\beta=40$, $\mu=-300\GeV$, $A=-100\GeV$, 
$m_{\tilde{Q}}= 300\GeV$, $M_A=500\GeV$ and $M_2=400\GeV$.
Focusing for $\sqrt{s}$ up to $1\TeV$,  the
various contributions to the tree level cross section
are presented (right).
The dotted-dashed line represents the  
tree-level cross section $\sigma_0^{WW}$, 
the dotted line the
$\sigma_0^{WW}+\sigma_0^{h-str}$. The dashed line 
includes also the interference term $\sigma_0^{\rm interf.}  $ and
represents the total tree-level cross section.
The solid line includes  the  one-loop correction.
} 
\label{fig:xsecs}
\end{figure}

We will start discussing the light Higgs boson production
$e^+ e^- \goes \bar{\nu} \nu h^0$, in the
MSSM, and the impact of the fermion/sfermion corrections calculated
in section~\ref{sec:loop}. 
In Fig.~\ref{fig:xsecs} (left) we 
have plotted the total tree-level cross section (dashed line) and
the one-loop corrected one (solid line) for $\sqrt{s}$ up to
$3\TeV$.
The tree-level cross section
includes the contribution from the $WW$ fusion channel, the Higgsstrahlung
and their interference. 
The  correction stemming from the fermion/sfermion
loops is always negative and substantial being of the order of $-10\%$.
Focusing for $\sqrt{s}$ up to $1\TeV$ (right) in a more detailed
figure,  the various contributions are presented.  
The dotted-dashed line represents the contribution from
the $WW$ channel at tree-level alone, whereas the dotted line includes
the Higgsstrahlung contribution as well. The dashed line
includes in addition the interference between the $WW$ channel and
Higgsstrahlung. The size of this
interference term is extremely small, hence
the difference between the dotted and dashed lines is rather tiny. 
It can also be seen  that for $\sqrt{s} \gtrsim 500 \GeV$ the
$WW$ fusion contribution  dominates the total cross section
for the Higgs production. Actually, for  
$\sqrt{s} \gtrsim 800 \GeV$  the total tree-level cross section
is due to  $WW$ fusion.
The  solid line represents again  the
one-loop corrected cross section.
In   Fig.~\ref{fig:xsecs} we
have taken: 
$\tan\beta=40$, $\mu=-300\GeV$,  $A=-100\GeV$, 
$ m_{\tilde{Q}}= 300\GeV$,
$M_A=500\GeV$, and $M_2=400\GeV$.
Choosing different sets of parameters,
the basic characteristics of these plots remain unchanged.  
Actually, the soft gaugino masses $M_{1,2}$ affect only the Higgs
boson masses and couplings through radiative corrections.

\begin{figure}[t]  
\begin{center}
\includegraphics[scale=.75]{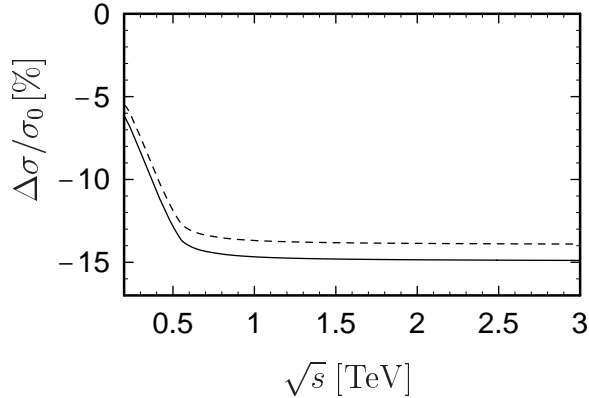} 
\end{center}

\caption[]{The relative correction $\Delta \sigma / \sigma_0$
as a function of $\sqrt{s}$ ($\Delta \sigma= \sigma - \sigma_0$),
where $\sigma_0$ is the tree-level and $\sigma$ the
one-loop corrected cross section.
The solid and dashed lines correspond to two different
choices of the SUSY parameters, as described in the  text.
} 
\label{fig:relcor}
\end{figure}

\begin{figure}[t]  
\begin{center}
\includegraphics[scale=.75]{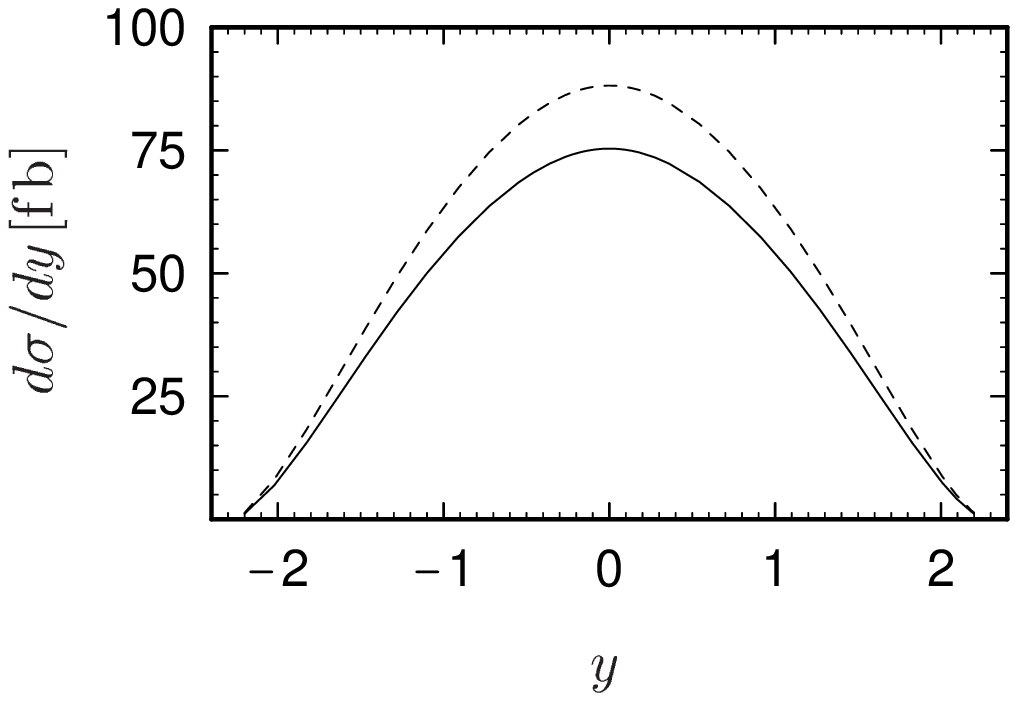}  \hspace*{2mm} 
\includegraphics[scale=.75]{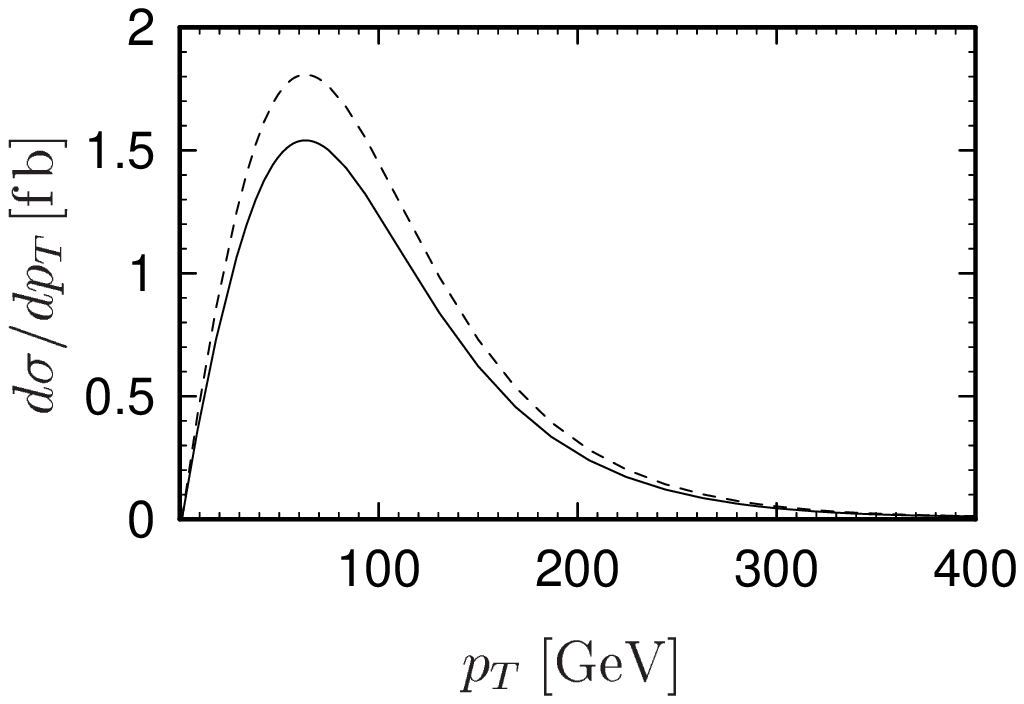} \\[1cm]
\includegraphics[scale=.75]{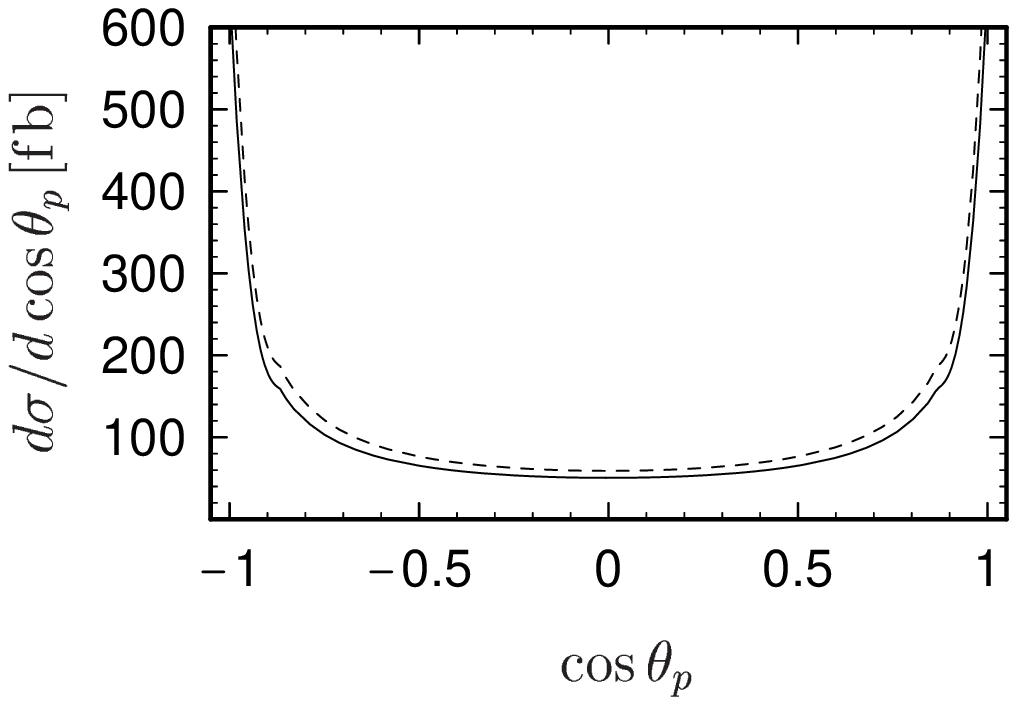} 
\end{center}

\caption[]{The distributions $\frac{d \sigma}{d y}$, 
$\frac{d \sigma}{d p_T}$ and $\frac{d \sigma}{d \cos\theta_p}$
as a function of the rapidity $y$, the transverse momentum $p_T$
and  $\cos\theta_p$, respectively. 
Here $\sqrt{s}=1\TeV$.
The dashed line (solid line) represents the tree-level (one-loop corrected) 
distribution. 
The SUSY parameters are as in Fig.~\ref{fig:xsecs}.
} 
\label{fig:ydis}
\end{figure}

In  Fig.~\ref{fig:relcor} the relative 
correction $\Delta \sigma / \sigma_0$ is presented as a
function of $\sqrt{s}$ for two different sets of parameters.
The solid line corresponds to the set used in  the Fig.~\ref{fig:xsecs},
 whereas for the
dashed line we have taken  
$\tan\beta=10$, $\mu=-100\GeV$ and $A=-500\GeV$, keeping
the rest of them unchanged.
This figure shows that the size of the one-loop correction to the
Higgs production cross section is practically
constant for $\sqrt{s} > 500 \GeV$ and weighs about $-15 \%$,
almost for any choice   of the SUSY parameters.
This  is a consequence of the dominance of the fermion-loop contribution
over  the one-loop corrections, and
therefore the total correction is not very sensitive
to the choice of the SUSY parameters.
This behaviour of the one-loop correction will be discussed
further after presenting the influence of the corrections
on the various distributions.

In Fig.~\ref{fig:ydis} we present the distributions  $\frac{d \sigma}{d y}$, 
$\frac{d \sigma}{d p_T}$ and $\frac{d \sigma}{d \cos\theta_p}$
as a function of the rapidity $y$, the transverse momentum $p_T$
and  $\cos\theta_p$, respectively. We have fixed $\sqrt{s}=1\TeV$.
The dashed lines represent the tree-level case, while the
solid lines the one-loop corrected one.
The SUSY parameters are as in Fig.~\ref{fig:xsecs}.
Paying attention to the figures of the  $\frac{d \sigma}{d y}$
and $\frac{d \sigma}{d \cos\theta_p}$ distributions, we
see that the tree-level distributions are completely symmetric.
We have  checked numerically that the one-loop 
corresponding distributions are also symmetric up to 
differences of $\mathcal{O}(10^{-2})$ fb.
For example, at a collider like
TESLA with an integrated luminosity of $500 \, {\rm fb}^{-1}$
such a small asymmetry yields only few events, making
these measurements very difficult.
The reason for such tiny asymmetries is the smallness of the
form factor $F^{21}$, which contributes to the one-loop corrections
in Eqs.~(\ref{eq:wwcor}) and (\ref{eq:zzcor}).  
The dominant one-loop correction results from the form factor $\hat{F}^{00}$,
which having the same structure as the tree-level $WWh^0$ coupling,
that is a correction to the $g^{\mu\nu}$ term in Eq.~(\ref{eq:lagr}),
does not contribute to the asymmetry.

This fact, in conjunction with  the behaviour of the correction
illustrated in Fig.~\ref{fig:relcor}, suggests
that a handy approximation of these fermion/sfermion loop corrections 
might be possible~\cite{EMSapp}.
The major complication in calculating the corrections from
 Eqs.~(\ref{eq:wwcor}) and (\ref{eq:zzcor}) is the dependence
of the form factors $\hat{F}^{00}$, $F^{21}$ on the
momentum transfer of the fused $W$-bosons $k_{1,2}^2$.
On the other hand, the dominant contribution from the
integration of  these terms over  the phase-space
arises for small values of $k_{1,2}^2$. Therefore, the essence of such
an approximation will be to keep the dominant Yukawa terms
from the form factor $\hat{F}^{00}$ for  $k_{1,2}^2 \sim 0$, which
will be just a factor correction to the tree-level coupling $WWh^0$.
In the literature there is an effective approximation,
where one only corrects the $WWh^0$ coupling~\cite{kniehl1}.
Although the sign of this approximation
is correct, it does not however account for the whole effect.

\begin{figure}[t]  
\begin{center}
\includegraphics[scale=.75]{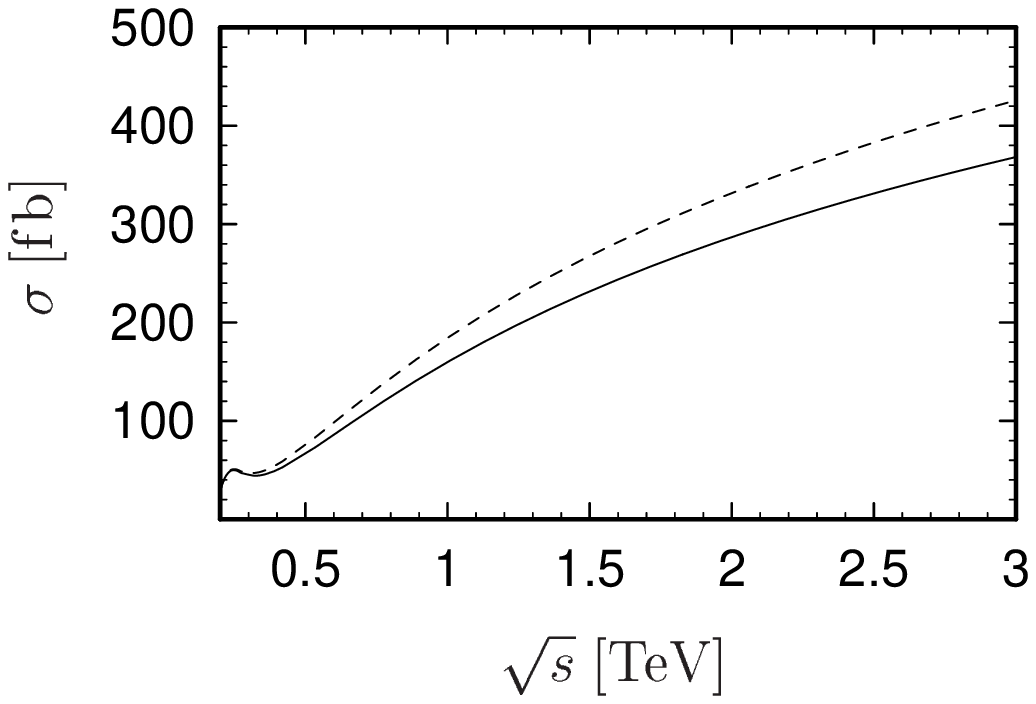} 
\includegraphics[scale=.75]{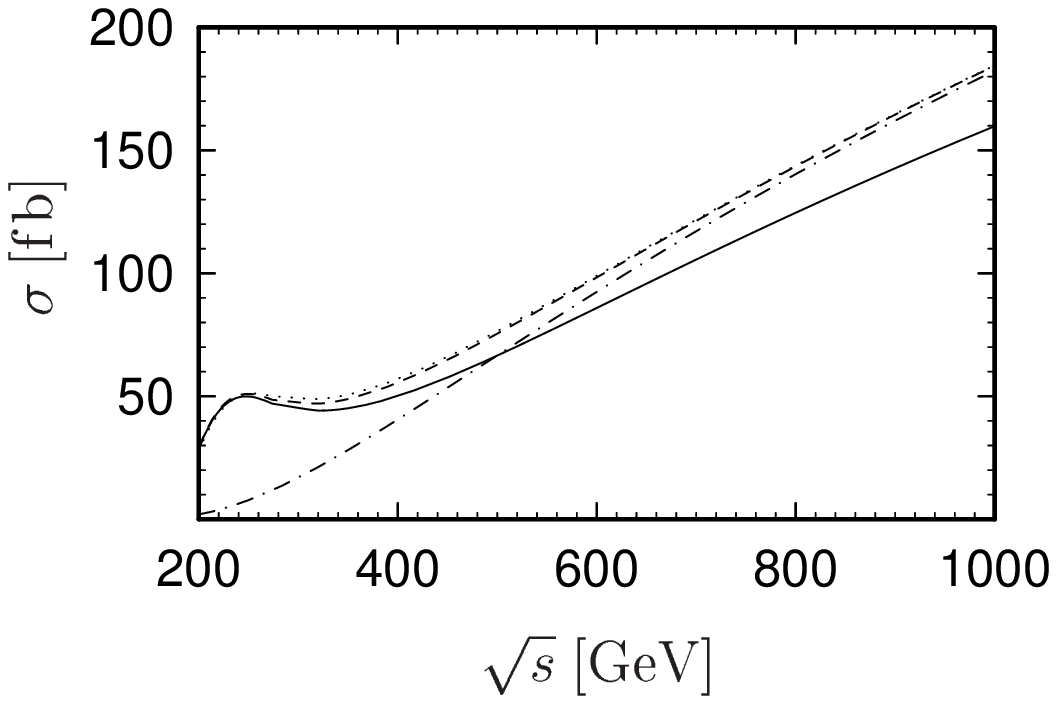}
\end{center}

\caption[]{The various cross sections for the heavy Higgs
boson $H^0$ production in the ``intense coupling regime''.
The meaning of the curves in the plots is as in Fig.~\ref{fig:xsecs}.
The SUSY parameters are chosen as
$\tan\beta=30$, $\mu=M_2=350\GeV$, $A=1000\GeV$, 
$m_{\tilde{Q}}= 1000\GeV$, $M_A=110\GeV$.
} 
\label{fig:xsecsH}
\end{figure}

As it has been discussed earlier, in the bulk of the SUSY parameter
space the  $WWH^0$ coupling is rather small, resulting in a
small production cross section for the heavy $CP$-even Higgs boson
$H^0$. The situation can be reversed in 
the so-called ``intense coupling regime''.
There the $WWH^0$ coupling can be significant, while the $WWh^0$ coupling
becomes smaller. For this case we  show Fig.~\ref{fig:xsecsH}.
In order to approach this case the SUSY parameters have been
chosen as $\tan\beta=30$, $\mu=M_2=350\GeV$, $A=1000\GeV$, 
$m_{\tilde{Q}}= 1000\GeV$, $M_A=110\GeV$. For such a choice
$h^0$, $H^0$ and $A^0$ are almost degenerate and light with
masses from $110\GeV$ to $120\GeV$. 
The meaning of the various curves in the plots here is as 
the corresponding in the  Fig.~\ref{fig:xsecs}.
Comparing  Fig.~\ref{fig:xsecsH} with  Fig.~\ref{fig:xsecs}
we see that the cross section for the $H^0$ production is smaller
than the $h^0$ production. Yet, it is possible to tune the
SUSY parameters in such a way to obtain values for the heavy
Higgs production as large as for the light one.
In any case, it seems that going to the ``intense coupling regime''
the task of discriminating between the two Higgs bosons 
becomes not trivial.

\begin{figure}[t]  
\begin{center}
{\setlength{\unitlength}{1mm}
\begin{picture}(160,55)(0,0)
\put(0,0){\includegraphics[scale=.75]{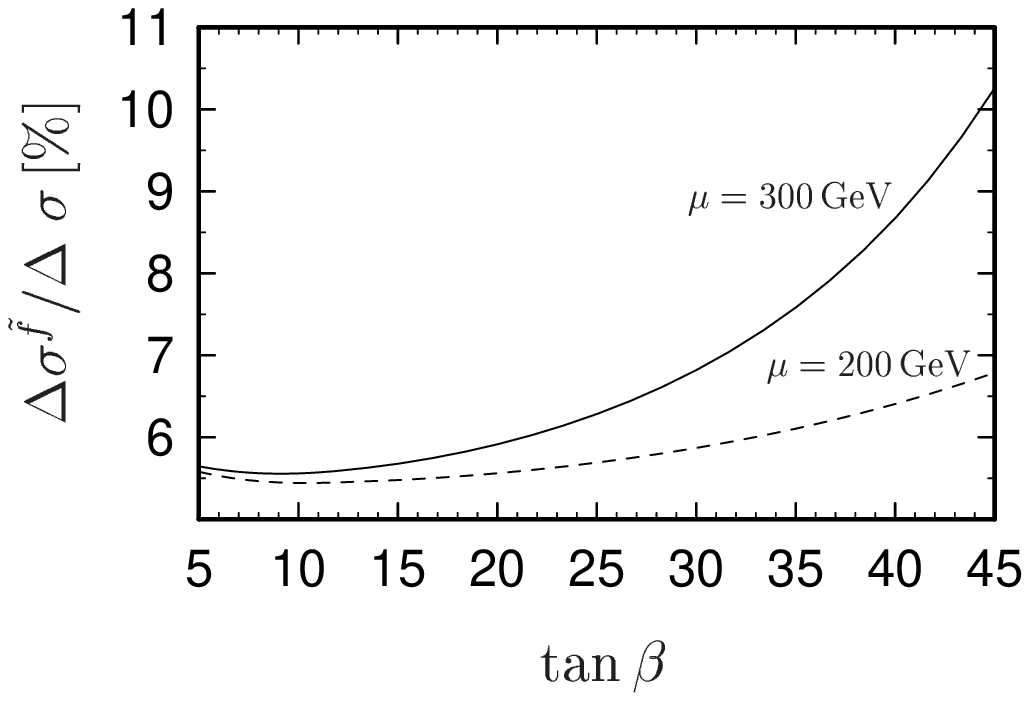}}
\put(80,-0.8){\includegraphics[scale=.75]{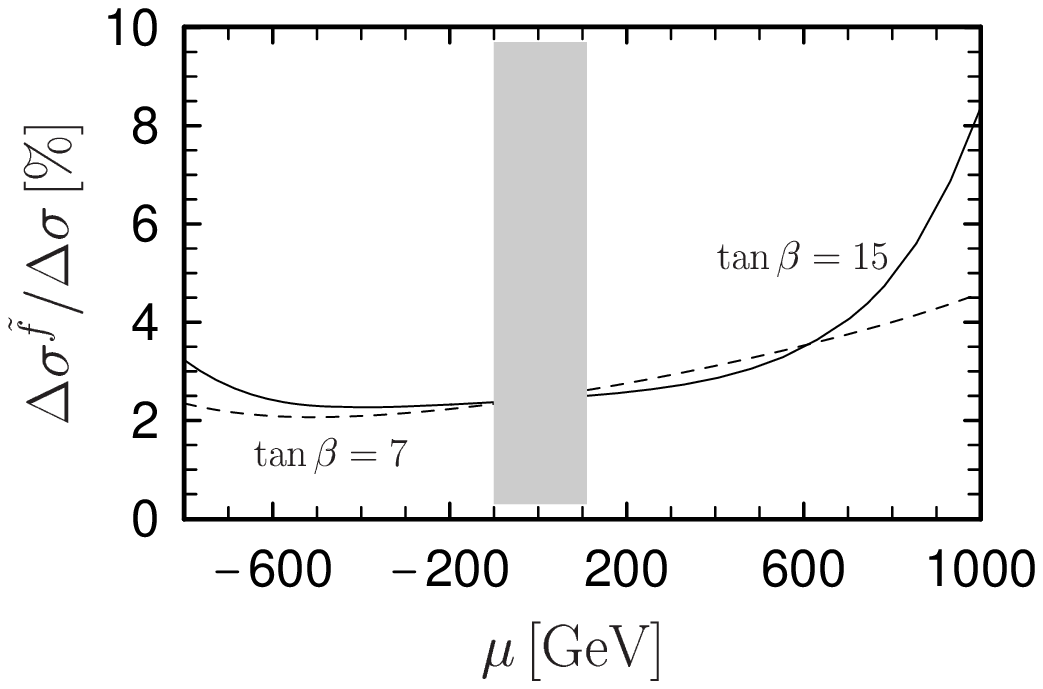}}
\end{picture}
} 
\end{center}

\caption[]{The percentage of the sfermions to the total one-loop
correction as a function of $\tan\beta$ (left) and  $\mu$ (right).
The rest of the SUSY parameters have been fixed as described in the
text. Here  $\sqrt{s}=1\TeV$.
The grey area in the right figure is excluded due
to the chargino mass bound.
}
\label{fig:susy}
\end{figure}

Fig.~\ref{fig:susy} exhibits the percentage of the
sfermion loops to the total one-loop correction as a function
of $\tan\beta$ (left) and $\mu$ (right), for
two different values of $\mu$ and $\tan\beta$, respectively, as
shown in the figure. 
In the left (right) figure we have chosen $A=100\GeV$ ($A=400\GeV$).
The rest of the SUSY parameters are:
$ m_{\tilde{Q}}= 300\GeV$, $M_A=500\GeV$, and $M_2=400\GeV$.
Here  $\sqrt{s}$ has been fixed to $1\TeV$.
The grey area in the right figure is excluded due
to the chargino mass bound.
 We see that the maximum value
of order 10\% can be achieved for large values of $\mu$ and
$\tan\beta$. There, due to the significant mixing in the stop and sbottom
sector, the  contribution of stops and 
sbottoms in the loops is enhanced. 
For these values of SUSY parameters the sfermion masses approach their
experimental lower bounds.
But even there the dominant correction, at least  $90\%$ of the 
total correction, is due to the fermion loops.

Finally, let us discuss  the background to single
Higgs boson production, Eq.~(\ref{eq:LCsingleHiggs}).
To the background several processes contribute. Single $Z$-boson 
production from 
$W W$ fusion has a large cross section ($\sim$ 200~fb at $\sqrt{s}$ = 500 GeV) 
and a similar topology as the signal process but the invariant mass of the two 
jets would peak at $m_{Z}$. Tagging of $b$ would improve the signal as the 
branching ratio of $Z \to b \bar b$ is only 15\%. The angular distribution
of the two jets would also be different due to the spin~1 of the $Z$ boson.
Double $Z$-boson production, 
$e^+ e^- \to Z Z$, where one $Z$ decays into $\nu \bar\nu$ 
and the other  into 
two jets ($\sigma \sim$ 500~fb at $\sqrt{s}$ = 500 GeV)
is another background. 
It can be reduced by cutting out the forward direction and 
measuring the invariant mass of the two jets. An important background is due
to the process $e^+ e^- \to e^+ W^- \nu_e$ through $\gamma W$ fusion 
($\sigma \sim$ 4.5 pb), 
with a very low $p_T$ electron being lost in the beam pipe.
Again the invariant mass of the two jets from $W$ would give a peak at $m_W$, 
and  above all $b$-tagging would 
strongly reduce this background \cite{desy92}.
Another source for the background is 
due to $e^+ e^- \to e^+ e^- b \bar b$ (via
$\gamma \gamma$ fusion), where the $e^+$ and $e^-$ are emitted in the very 
forward direction thereby escaping detection. 
However, the significance of this 
background can be substantially reduced by 
making a cut in $|\cos\theta|$ of the 
outgoing $b \bar b$ pair, to eliminate the part, where the $b \bar b$ is
emitted near the beam direction \cite{linac92}.

\section{Conclusions}
\vspace{-3mm}
\label{sec:concl}
In this paper we have calculated the fermion/sfermion loop
corrections to  single Higgs boson production $\higproo$ in
the context of the MSSM. 
They are supposed to be the dominant radiative corrections due
to the size of the Yukawa couplings.
At the next generation of high energy $e^+ e^-$ linear colliders, 
where $\sqrt{s} \gtrsim 500 \GeV$ , the $WW$ fusion
channel dominates the Higgs boson production cross section. 
We have also  included the Higgsstrahlung process and its interference
with the $WW$ fusion.
The  one-loop correction to the cross section 
is negative and of the order of $-10\%$,
and is rather independent of  $\sqrt{s}$ for $\sqrt{s} > 500 \GeV$.
It is dominated  by the fermion loops, usually being larger than 
$90\%$ of the total correction.
For the case of  maximal mixing in the sfermion mass matrices,
the contribution of the sfermion loops is enhanced, but
nevertheless weighs less than $10\%$ of the total one-loop correction.
The possibility of having polarized  $e^+ / e^-$ beams is also explored.
The study of the kinimatical distributions of the rapidity and the 
production angle of the Higgs boson shows
that these loop corrections do not alter the symmetry of the
tree-level distributions. 
In the bulk of the parameter space of the MSSM the $WWH^0$ coupling 
is suppressed, making the heavy Higgs boson production very difficult.
Yet, going to the ``intense coupling regime'' there is
a possibility of obtaining sizable values for this coupling.
We have studied the heavy Higgs boson production, including the  one-loop
fermion/sfermion corrections, in this case.


\vfill
\noindent 
{\bf Note added in proof} \\ 
\noindent 
After submitting our paper, it was claimed in the paper by T.~Hahn, S.~Heinemeyer,
and G.~Weiglein, {\tt hep-ph/0211204}, that there is a discrepancy between their
and our results. The difference, however, was resolved in a recent paper by
A.~Denner et al., {\tt hep-ph/0301189}. The difference in the size of the 
radiative corrections is due to the use of a different charge renormalization 
scheme and different input parameters. With the same input mass parameters we get
the same result for the corrected cross section.\\
We thank A.~Denner and T.~Hahn et al. for comparing results. Especially, we 
want to thank A.~Denner for correspondence and the strong effort in clarifying 
the situation. 

\vfill
\noindent 
{\bf Acknowledgements} \\ 
\noindent 
V.~C.~S. acknowledges support by a Marie Curie Fellowship of the EU
programme IHP under contract HPMFCT-2000-00675.
The authors acknowledge support from
EU under the HPRN-CT-2000-00149 network programme.
The work was also supported by the ``Fonds zur F\"orderung der
wissenschaftlichen Forschung'' of Austria, project No. P13139-PHY.

\newpage
\appendix
\noindent{\bf{\Large Appendices}}
\section{Form factors \boldmath $F^{00}$ and $F^{21}$}
\label{sec:app_ff}
For convenience we are presenting the formulae for the (s)top--(s)bottom
doublet, but actually they are valid for any (s)fermion doublet.
The form factors  $F^{00}$ and $F^{21}$ are given by
\begin{eqnarray}
  F^{00} & = & \frac{1}{(4 \pi)^2} \left(A^{00}_{btt} + A^{00}_{tbb} + A^{00}_{3 \sq} +
A^{00}_{2 \sq}\right)\, ,
\label{eq:formfac00}\\
  F^{21} & = & \frac{1}{(4 \pi)^2} \left( 
A^{21}_{btt} + A^{21}_{tbb} + A^{21}_{3 \sq}\right)\, .
\label{eq:formfac21}
\end{eqnarray}
Note that $A^{21}_{2 \sq} = 0$.
$A^{00}_{btt}$, $A^{21}_{btt}$, $A^{00}_{tbb}$, and $A^{21}_{tbb}$ 
correspond to Fig.~\ref{fig:feyn}a, 
$A^{00}_{3 \sq}$ and $A^{21}_{3 \sq}$ to Fig.~\ref{fig:feyn}b, and
$A^{00}_{2 \sq}$ to Fig.~\ref{fig:feyn}c.
\begin{eqnarray}
  A^{00}_{btt} & = & \frac{g^3\, m_t^2}{4\, m_W\, \sin\beta}\,
\{\cos\alpha,\, \sin\alpha\}\,
\Big((2\,m_t^2 + 2\, m_b^2 - k_1^2 - k_2^2)\, C_0 - 8\, C_{00}\nn\\
&&\hspace{2.5cm} + B_0(k_1^2, m_t^2, m_b^2) + B_0(k_2^2, m_t^2, m_b^2) 
+ 2\, B_0(m_{H^0_k}^2, m_t^2, m_t^2) \Big) 
 \, ,
\label{eq:A00btt}\\
  A^{21}_{btt} & = &  \frac{g^3\, m_t^2}{2\, m_W\, \sin\beta}\,
\{\cos\alpha,\, \sin\alpha\}\,
\Big(C_0 + C_1 + 4 (C_2 + C_{12} + C_{22}) \Big) \, ,
\label{eq:A21btt}
\end{eqnarray}
with $C_{..} =
C_{..}(k_2^2, m_{H^0_k}^2, k_1^2, m_b^2, m_t^2, m_t^2)$, and
$\{\cos\alpha,\, \sin\alpha\}$ corresponds to $k = \{1,\, 2\}$.
\begin{eqnarray}
  A^{00}_{tbb} & = & \frac{g^3\, m_b^2}{4\, m_W\, \cos\beta}\,
\{-\sin\alpha,\, \cos\alpha\}\,
\Big((2\,m_t^2 + 2\,m_b^2 - k_1^2 - k_2^2)\, C_0 - 8\, C_{00}\nn\\
&&\hspace{2.5cm} + B_0(k_1^2, m_t^2, m_b^2) + B_0(k_2^2, m_t^2, m_b^2) 
+ 2\, B_0(m_{H^0_k}^2, m_b^2, m_b^2) \Big)
 \, ,
\label{eq:A00tbb}\\
  A^{21}_{tbb} & = &  \frac{g^3\, m_b^2}{2\, m_W\, \cos\beta}\,
\{-\sin\alpha,\, \cos\alpha\}\,
\Big(C_0 + C_1 + 4 (C_2 + C_{12} + C_{22}) \Big) \, ,
\label{eq:A21tbb}
\end{eqnarray}
with $C_{..} =
C_{..}(k_2^2, m_{H^0_k}^2, k_1^2, m_t^2, m_b^2, m_b^2)$, and
$\{-\sin\alpha,\, \cos\alpha\}$ corresponds to $k = \{1,\, 2\}$.
\begin{equation}
A^{ab}_{3 \sq} = -2 \, g^2 \sum_{i,j,l = 1,2}\,
\Big(G_{ijk}^\st  R^\st_{i1} R^\st_{j1} R^\sb_{l1}
R^\sb_{l1}\, C^1_{ba} + G_{ijk}^\sb  R^\sb_{i1} R^\sb_{j1} R^\st_{l1}
R^\st_{l1}\, C^2_{ba} \Big)\, ,
\label{eq:A3sq}
\end{equation}
with the couple of indices $(a b)$ = $(0 0)$ or $(2 1)$,
$k = 1,2$ denotes $h^0$ and $H^0$, $C^1_{..} =
C_{..}(k_2^2, m_{H^0_k}^2, k_1^2, m_{\sb_l}^2, m_{\st_i}^2, m_{\st_j}^2)$
and $C^2_{..} =
C_{..}(k_2^2, m_{H^0_k}^2, k_1^2, m_{\st_l}^2, m_{\sb_i}^2, m_{\sb_j}^2)$.
The definition of the rotation matrices $R^\st$ and $R^\sb$,
and the coupling matrices
$G_{ijk}^\st$ and $G_{ijk}^\sb$ are given in Ref.~\cite{EMKY}. 
\begin{equation}
A^{00}_{2 \sq} = \frac{g^2}{2} \sum_{i,j = 1,2}\,
\Big( G_{ijk}^\st  R^\st_{i1} R^\st_{j1} 
       B_0(m_{H^0_k}^2, m_{\st_i}^2, m_{\st_j}^2) 
+ G_{ijk}^\sb  R^\sb_{i1} R^\sb_{j1} 
         B_0(m_{H^0_k}^2, m_{\sb_i}^2, m_{\sb_j}^2)\Big)\, .
\label{eq:A2sq}
\end{equation}

\section{Calculation of the cross section}
\label{sec:app_xsec}
In general the differential cross section 
for the process of Eq.~(\ref{eq:LCsingleHiggs}), 
when the colliding fermions $f_1$, $f_2$ are massless, is given by
\beq
d \sigma=
 \frac{ |\mathcal{M}|^2}{4 \,p_1\cdot p_2}\;
 d\Phi_3 \,,
\eeq 
where the 3-body phase-space is
\beq
d\Phi_3 =(2 \pi)^4 \; \delta^4(p_1 + p_2 - p_3 - p_4 -p) \;
     \frac{d^3 p_3}{(2\pi)^3\, 2E_3} \;
       \frac{d^3 p_4}{(2\pi)^3\, 2E_4} \;
       \frac{d^3 p}{(2\pi)^3\, 2E_p} \,.
\eeq
The differential cross section can be cast into the form
\beq
E_p \, \frac{d^3 \sigma}{d^3 p}=
        \int \frac{|{\mathcal{M}}|^2}{16 \, s \, (2\pi)^5}
  \;  \delta^4(p_1 + p_2 - p_3 - p_4 -p) \; 
\frac{d^3p_3}{E_3} \; \frac{d^3p_4}{E_4} \; ,
\label{eq:gxsec}
\eeq
where $s=(p_1+p_2)^2=2 \, p_1 \cdot p_2 $.

In order to calculate the differential cross section of Eq.~(\ref{eq:gxsec})
we are following the procedure of Ref.~\cite{alta} and
we choose to work in the rest  frame of the two final fermions defined 
by $\vec{p}_3 +\vec{p}_4 =0 $, see Fig~\ref{frame34}.
In this frame $\vec{p}=\vec{p}_1 + \vec{p}_2 $, which
means that the vectors $\vec{p}$, $\vec{p}_1$ and
$\vec{p}_2$ lie in the same plane, the $(x,z)$ plane in our case.

In this frame one gets
\bea
E_p \, \frac{d^3 \sigma}{d^3 p} &=&
               \int \frac{|{\mathcal{M}}|^2}{16\,s\, (2\pi)^5}
  \,\,  \delta(E_1 + E_2 - 2\, E_3  -E_p) \,
   d E_3 \,\,  d \cos\theta \, \,d \phi  \nonumber \\ 
 &=& \int_{-1}^{1}\, d \cos\theta \int_{0}^{2\pi} \, d \phi \,
 \frac{|{\mathcal{M}}|^2}{ s \, (4\pi)^5} \, .
\label{eq:xsec1}
\eea
The  products $p_i \cdot p_j$, $i,j=1,2,3,4$, which
are involved in $|\mathcal{M}|^2$,  
can be expressed in terms of  the angles $\theta$, $\phi$, $\chi$
and the products $p \cdot p_1$, $p \cdot p_2$ :
\bea
p_1 \cdot p_2 &=& \frac{s}{2} \,,\nonumber \\ 
p_1 \cdot p_3 &=& {\textstyle{\frac{1}{4}}}\, 
                 (s-2 \, p\cdot p_1)\,(1-\cos\theta)\,,\nonumber \\ 
p_1 \cdot p_4 &=& {\textstyle{\frac{1}{4}}} \, 
                     (s-2\, p\cdot p_1)\,(1+\cos\theta)\,,\nonumber \\ 
p_2 \cdot p_3 &=& {\textstyle{\frac{1}{4}}} \, (s-2\, p\cdot p_2)\,
    (1 -\cos\chi \cos\theta -\sin\chi \sin\theta \cos\phi )\,,\nonumber \\
p_2 \cdot p_4 &=& {\textstyle{\frac{1}{4}}} \, (s-2\, p\cdot p_2)\,
    (1 +\cos\chi \cos\theta +\sin\chi \sin\theta \cos\phi )\,,\nonumber \\
p_3 \cdot p_4 &=& {\textstyle{\frac{1}{2}}}\,
            (s+m_H^2-2 \, p\cdot p_1 -2 \, p\cdot p_2)\,,\nonumber \\
k_1^2&=&-2\, p_1 \cdot p_3 \,,\nonumber \\
k_2^2&=&-2\, p_2 \cdot p_4 \,.
\label{sys}
\eea 
It is important to note that 
$\cos\chi$ can be  expressed in terms of
Lorentz invariant quantities
\beq
\cos\chi=1- \frac{2\,s\,(s+m_H^2-2 \, p\cdot p_1 -2 \, p\cdot p_2)}
            {(s-2 \, p\cdot p_1)\, (s-2 \, p\cdot p_2)} \,. 
\label{cchi}
\eeq
$m_H$ denotes the mass of the Higgs boson $h^0/H^0$.

Using Eqs.~(\ref{sys}),(\ref{cchi}) and integrating over the angles
$\phi$ and $\theta$ in Eq.~(\ref{eq:xsec1}) one obtains the
differential cross section $ E_p \frac{d^3 \sigma}{d^3 p} $
as a function of the Lorentz invariants: $\cos\chi$,
$ p\cdot p_1 $ and $ p\cdot p_2 $.
In order to calculate the total cross section we go to the rest frame
of the initial fermions. 
There we have $\vec{p}_1 + \vec{p}_2 =0$, see Fig.~\ref{frame12},
and we have chosen the beam direction as the $z$-axis. 
The three momenta of the final state particles $\vec{p}$,
$\vec{p}_3$ and $\vec{p}_4$ span a plane, and 
$\theta_p$ is the angle between the $\vec{p}_1$ and  $\vec{p}$.
In this reference frame one finds
\bea
&& p \cdot p_1 = \frac{\sqrt{s}}{2}
              (E_p - |\vec{p} \, |\, \cos\theta_p) \;\;,\;\;
   p \cdot p_2 = \frac{\sqrt{s}}{2
              }(E_p + |\vec{p}\, | \, \cos\theta_p) \,,\nonumber \\ 
&& |\vec{p} \, | = \sqrt{E_p^2 - m_H^2} \,.
\label{sys2}
\eea
The total cross section is given by
\bea
\sigma &=& \int_0^{2\pi} d\phi' \int_{-1}^1 d \cos\theta_p 
\int_0^{|\vec{p}\, |^{\mathrm{max}}} d |\vec{p}\, |  \,\, |\vec{p}\, |^2
\,\, \frac{1}{E_p}\,\left(E_p \, \frac{d^3 \sigma}{d^3 p} \right) \nonumber \\ 
&=& 2 \pi \, \int_{-1}^1 d \cos\theta_p \,\,
\int_{m_H}^{E_p^{\mathrm{max}}} dE_p \,\,
  |\vec{p}\, | \left(  E_p \, \frac{d^3 \sigma}{d^3 p} \right) \,.
\label{eq:xsec2}
\eea
The maximum value of  $|\vec{p} \, |$
\beq
|\vec{p}\,|^{\mathrm{max}}=\frac{s-m_H^2}{2\, \sqrt{s}} 
\label{pmax}
\eeq 
is for $\cos\delta =1 $, see Fig.~\ref{frame12}, and
this gives
\beq
E_p^{\mathrm{max}}=\frac{s+m_H^2}{2\, \sqrt{s}} 
\eeq 
as integration limit for Eq.~(\ref{eq:xsec2}).

It is more convenient  to use  in Eq.~(\ref{eq:xsec2}) the
rapidity $y$ and the transverse momentum
$p_T$ of the produced Higgs boson,  instead of 
the integration variables $E_p$ and $\cos\theta_p$.  
The transverse and longitudinal momenta of the Higgs boson are
\beq
p_T=|\vec{p}\,| \sin\theta_p \,\, , \,\, 
p_L=|\vec{p}\,| \cos\theta_p \,. 
\eeq
Defining the rapidity as
\beq
y = \frac{1}{2} \ln \left( \frac{E_p+p_L}{E_p-p_L} \right) \, ,
\eeq 
one finds
\bea
&& p_L = m_T \, \sinh y \;\; , \;\;  
   p \cdot p_1 = \frac{\sqrt{s}}{2}\,\, m_T \,\,e^{-y}\,, \nonumber \\
&& E_p = m_T \, \cosh y   \;\; , \;\;
   p \cdot p_2 = \frac{\sqrt{s}}{2}\,\, m_T \,\, e^{y}\,\,,
\eea
where the transverse mass of the Higgs boson
is defined as
\beq
m_T \equiv \sqrt{m_H^2 + p_T^2} \,.
\eeq

Changing the integration variables from $\cos\theta_p,E_p$
to  $y,p_T$, the total cross section of Eq.~(\ref{eq:xsec2}) can 
be cast into the form
\beq
\sigma = \int_{y_{-}}^{y_{+}} dy 
         \int_0^{(p_T^2)^{\mathrm{max}}}d p_T^2 
     \left( \frac{d^2 \sigma} {dy \, dp_T^2} \right) \,,
\label{eq:xsec3}
\eeq 
where
\beq
\left( \frac{d^2 \sigma}{dy \, dp_T^2} \right) =
\pi \left(E_p \, \frac{d^3 \sigma}{d^3 p} \right) \,.
\label{eq:relation}
\eeq
For the integration limits we find
\beq
y_{\pm}=\pm \ln \frac{\sqrt{s}}{m_H} \,\, , \,\, 
(p_T^2)^{\mathrm{max}}=
\left( \frac{s+m_H^2}{2\, \sqrt{s} \, \cosh y} \right)^2 - m_H^2 \,.
\label{eq:limits}
\eeq
One can reverse the order of the integrations and to
carry out first the the integration over the 
rapidity $y$. In order to do this,
we have to inverse the function 
$(p_T^2)^{\mathrm{max}}(y)$ in Eq.~(\ref{eq:limits}).
By doing this and by studying the integrations limits in
the plane ($y,p_T^2$) one gets
\beq
\sigma =   \int_0^{(\tilde{p}_T^2)^{\mathrm{max}}}   d p_T^2 
           \int_{\tilde{y}_{-}}^{\tilde{y}_{+}} dy 
           \left( \frac{d^2 \sigma} {dy \, dp_T^2} \right) \,.
\label{eq:xsec4}
\eeq 
For the new integration limits we find
\beq
 (\tilde{p}_T^2)^{\mathrm{max}} =
   \left( \frac{s-m_H^2}{2 \, \sqrt{s} } \right)^2  \,,\,\,
\tilde{y}_{\pm}=\pm \ln \left[z+\sqrt{z^2-1}\right] \,,
\eeq
where  
\beq
z=\frac{s+m_H^2}{2\, \sqrt{s\, (p_T^2+m_H^2)}} \, \,.
\eeq


\clearpage


\newpage
\begin{thebibliography}{99} 

\bibitem{all}
See for example, Particle Data Group, K.~Hagiwara \etal, 
Phys. Rev. D66 (2002) 010001.

\bibitem{higgs}
LEP Collaborations, CERN-EP/2001-055 (2001), hep-ex/0107021;
LEP Higgs Working Group, hep-ex/0107030,
{\tt http://lephiggs.web.cern.ch/LEPHIGGS}.

\bibitem{higgslim}
M. Carena, J.R. Espinosa, M. Quiros and C.E.M. Wagner,
Phys. Lett.  B355 (1995) 209;  M. Carena, M. Quiros and
C.E.M. Wagner, Nucl. Phys. B461 (1996) 407;
H.E. Haber, R. Hempfling and A.H. Hoang,
Z. Phys.  C75 (1997) 539;
S.~Heinemeyer, W.~Hollik and G.~Weiglein, Phys. Rev. D58 (1998)
091701; Phys. Lett. B440 (1998) 296; Eur. Phys. J.
C9 (1999) 343; Phys. Lett. B455 (1999) 179;
M. Carena, H.E. Haber, S. Heinemeyer, W. Hollik,
C.E.M. Wagner and G. Weiglein, Nucl. Phys. B580 (2000) 29;
J.R.~Espinosa and R.-J. Zhang, JHEP 0003 (2000) 026;
Nucl. Phys. B586 (2000) 3;
J.R.~Espinosa and I.~Navarro, Nucl. Phys. B615 (2001) 82;
G.~Degrassi, P.~Slavich and F.~Zwirner,
Nucl. Phys. B611 (2001) 403;
A.~Brignole, G.~Degrassi, P.~Slavich and F.~Zwirner,
Nucl. Phys. B631 (2002) 195; hep-ph/0206101.



\bibitem{tevatron}
Report of the Tevatron Higgs Working Group,
M.~Carena, J.~S.~Conway, H.~E.~Haber, J.~D.~Hobbs,
FERMILAB-Conf. 00/279-T, hep-ph/0010338.



\bibitem{carhaber}
M.~Carena, H.~E.~Haber, hep-ph/0208209.



\bibitem{aachen}
Proceedings of the Large Hadron Collider Workshop, Aachen 1990,
CERN 90-10, Vol. 1, ed. by G.~Jarlskog, D.~Rein.


\bibitem{zepe}
D.~Rainwater, D.~Zeppenfeld, JHEP 12 (1997) 5;
D.~Rainwater, D.~Zeppenfeld, K.~Hagiwara, Phys. Rev. D59 (1999) 014037;
T.~Plehn, D.~Rainwater, D.~Zeppenfeld, Phys. Lett. B454 (1999) 297. 

\bibitem{roeck}
A. De Roeck \etal, hep-ph/0207042.


\bibitem{jlc}
N.~Akasaka \etal, ``JLC design study,'' KEK-REPORT-97-1;
also presentation by K.~Yokoya, 
{\tt http://lcdev.kek.jp/Reviews/LCPAC2002/LCPAC2002.KY.pdf}.

\bibitem{nlc}
C.~Adolphsen \etal,  International Study Group Collaboration,
``International study group progress report on linear collider development'',
SLAC-R-559 and KEK-REPORT-2000-7.

\bibitem{tesla}
R.~Brinkmann, K.~Flottmann, J.~Rossbach, P.~Schmuser, N.~Walker and 
H.~Weise (editors),
``TESLA: The superconducting electron positron linear collider with an
integrated X-ray laser laboratory.  Technical design report, Part~2: The
Accelerator,'' DESY-01-011, {\tt http://tesla.desy.de/tdr/}.


\bibitem{clic}
R.W.~Assmann \etal, The CLIC Study Team,
``A 3~TeV $e^+e^-$ linear collider based on CLIC technology'', 
SLAC-REPRINT-2000-096 and CERN-2000-008, ed. by
G. Guignard.



\bibitem{fusion}
D.~R.~T.~Jones, S.~T.~Petcov, Phys. Lett. B84 (1979) 440;
R.~N.~Cahn, S.~Dawson, Phys. Lett. B136 (1984) 96;
G.~L.~Kane, W.~W.~Repko, W.~B.~Rolnick, Phys. Lett. B148 (1984) 367;
R.~N.~Cahn, Nucl. Phys. B255 (1985) 341;
B.~A.~Kniehl, Z. Phys. C55 (1992) 605.

\bibitem{alta} 
G.~Altarelli, B.~Mele, F.~Pitolli, Nucl. Phys. B287 (1987) 205.


\bibitem{kilian}
W.~Kilian, M.~Kr\"amer, P.~M.~Zerwas, 
Phys. Lett. B373 (1996) 135.


\bibitem{ellis}
J.~Ellis, M.~K.~Gaillard, D.~V.~Nanopoulos, 
Nucl. Phys. B106 (1976) 292;
J.~D.~Bjorken, Proc. Summer Institute on Particle Physics,
SLAC Report 198 (1976); B.~W.~Lee, C.~Quigg, H.~B.~Thacker, 
Phys. Rev. D16 (1977) 1519; B.~L.~Ioffe, V.~A.~Khoze, Sov. J. Part.
Nucl. 9 (1978) 50.  

\bibitem{kniehl-review}
B.~A.~Kniehl, Phys.~Rep.~240 (1994) 211.

\bibitem{kniehl1}
B.~A.~Kniehl, Phys. Rev. D53 (1996) 6477.

\bibitem{kniehl2}
B.~A.~Kniehl, M.~Steinhauser, Nucl. Phys. B454 (1995) 485.


\bibitem{hollik}
S.~Heinemeyer, W.~Hollik, J.~Rosiek, G.~Weiglein, 
Eur. Phys. J. C19 (2001) 535.


\bibitem{EMS}
H.~Eberl, W.~Majerotto, V.~C.~Spanos, Phys. Lett. B583 (2002) 353.


\bibitem{intense}
E.~Boos, A.~Djouadi, M.~M\"uhlleitner, A.~Vologdin, hep-ph/0205160.



\bibitem{EMKY}
H.~Eberl, W.~Majerotto, M.~Kincel, Y.~Yamada, Phys. Rev. D64 (2001) 115013;
Nucl. Phys. B625 (2002) 372. 


\bibitem{carena}
M.~Carena, M.~Quiros, C.~E.~M.~Wagner, Nucl. Phys. B461 (1996) 407.


\bibitem{EMSapp}
H.~Eberl, W.~Majerotto, V.~C.~Spanos, in preparation.


\bibitem{desy92}
P. Grosse Wiesmann, D. Haidt, H. J. Schreiber, 
$e^+ e^-$ Collisions at 500~GeV: The Physics Potential,
Part~A, p. 39, DESY 92-123A, ed. by P.~M.~Zerwas.

\bibitem{linac92}
H. E. Haber, Physics and Experiments with Linear Colliders,
Eds.: R. Orava, P. Eerola, M. Nordberg, World Scientific, Vol. 1,
p.~235, 1992.

 
\end{thebibliography}
\end{document}